\documentclass[a4paper,10]{article}
\usepackage{amsfonts}
\usepackage{amsthm}
\begin{document}
\title{Quantum Mechanics Without The Quantum}
\author{Constantin Antonopoulos [sections 1,2] and Theodossios Papadimitropoulos [section 3]}
\maketitle
\large
\begin{abstract}
\large The two Heisenberg Uncertainties entail an
\emph{incompatibility} between the two pairs of conjugated
variables $E,t$ and $p,q$. But incompatibility comes in \emph{two}
kinds, exclusive of one another. There is incompatibility
defineable as: $(p \rightarrow\neg q) \wedge (q \rightarrow\neg
p)$ \textbf{or} defineable as $[(p \rightarrow\neg q) \wedge (q
\rightarrow\neg p)] \Leftrightarrow r$. The former kind is
\emph{un}conditional, the latter conditional. The former, in
accordance, is fact \emph{in}dependent, and thus ascertainable by
virtue of \emph{logic}, the latter fact dependent, and thus
ascertainable by virtue of fact. The two types are therefore
diametrically opposed.

In spite of this, however, the existing derivations of the
Uncertainties are shown here to entail both types of
incompatibility simultaneously. $\Delta E \Delta t \geq h$, for
example, is known to derive from the quantum relation $E=h\nu$
plus the Fourier relation $\Delta\nu\Delta t \simeq 1$. And the
Fourier relation assignes a \emph{logical} incompatibility between
a $\Delta\nu=0$, $\Delta t=0$. (No frequency defineable at an
instant.) Which is therefore fact independent and unconditional.
How can one reconcile this with the fact that $\Delta E \Delta t$
\emph{if and only if}  $h>0$, which latter supposition is a
\emph{factual} truth, entailing that a $\Delta E=0$, $\Delta t=0$
incompatibility should itself be fact dependent?

To then say that the incompatibility at hand is only logical, i.e.
that resulting from $\Delta\nu\Delta t \geq 1$, is to treat
$\Delta E =0$, $\Delta t=0$ as \emph{un}conditionally
incompatible, since this is what their equivalents, $\Delta\nu=0$,
$\Delta t=0$ are, and therefore as incompatible independently of
the \emph{quantum}. And to say that it is only factual, amounts to
disputing $E=h\nu$ itself, whose presence alone is what
necessitates application of the -logical- relation
$\Delta\nu\Delta t \geq 1$. Since either option sacrifices an
equally essential requirement, it can only follow that this
Uncertainty expresses both a conditional \emph{and} an
unconditional form of incompatibility.

We continue by tracing the exact same phenomenon right within the
heart of the noncommutative formalism of QM. The \emph{fact}
dependent $p$,$q$ noncommutativity, expressed in $pq \neq qp$ as
derived from the relation $pq-qp=i \hbar I$, has given its place
to the abstract Hilbertian, fact independent noncommutativity,
expressed in $AB \neq BA$, without explicit or implicit reference
to $\hbar$. Hence, to identify the two would lead to a
contradiction comparable to the previous.
\end{abstract}
\section{Distinguishing Within Incompatibility}
In a series of previous works one of us \cite{AN88,AN94,AN97} has
argued that the incompatibility of the two pairs of conjugated
variables, comprising the action products $Et$ and $pq$, as
manifested in the two corresponding quantum uncertainties (UR here
after), comes in two, \emph{antithetic} types, once because this
is how theorists as a rule tend to argue (unfortunately), twice
because this is an option contained in the nature of
incompatibility itself. Indeed, and contrary to appearances or
common opinion, Incompatibility as such is a \emph{twofold}
concept. It is not too difficult to establish this in formal
logic. The dichotomy can be immediately seen (and felt) in the
following way:
\begin{itemize}
\item[a.]$(p \rightarrow\neg q) \wedge (q \rightarrow\neg p)$ however, also \item[b.]$[(p \rightarrow\neg q) \wedge (q \rightarrow\neg p)]
\Leftrightarrow r$.
\end{itemize}
These two expressions of Incompatibility are contradictory to one
another. For the possible value ascription $\neg r$ in [b] we will
obtain ``$p$ \emph{and} $q$'', a possibility which will never come
up within the contents of formula [a]. For ascription $\neg r$ to
[b], formulae [a] and [b] immediately develop incompatible truth
tables. This is because formula [a] expresses unconditional
incompatibility between p and q, while formula [b] only a
conditional incompatibility between them, conditional, to be
exact, on $r$. [b] reads: ``$p$ excludes $q$ and $q$ excludes $p$
if and only if $r$.'' But not otherwise. Since the possibility
that $\neg r$ stands for ``otherwise'', for the ascription $\neg
r$ the two propositional variables, $p$ and $q$, will cease to be
incompatible. But in formula [a] they never cease to be, come what
may.

In other words, the incompatibility between the two propositional
variables expressed in formula [b], as being conditional on the
presence of an additional factor (r), is one which obtains only in
\emph{some} cases. Namely, iff $r$. Since, however, the
incompatibility expressed in formula [a] is not conditional on
anything, this latter obtains independently of all other factors
whatosoever and hence obtains for \emph{all possible} cases
instead. It therefore (trivially) follows that no two pairs of
concepts can be ever both, conditionally as well as
unconditionally incompatible for no two pairs of concepts can ever
be both, incompatible in all possible cases and also incompatible
only in some.

The problem in QM is that, once this distinction is explicitly
drawn as above (which it never is), we frequently find ourselves
oblidged to conclude that the classical concepts, $E$ with $t$ and
$p$ with $q$, are indeed both. But of this later.

At present our task is to determine why and how -viz. under what
specific conditions Incompatibility presents itself in two
antithetic ways. To put the point differently, it may be clear to
us how the concepts ``one'' and ``many'' may be incompatible to
one another. They are as a matter of \emph{definition}. Hence, the
incompatibility between ``one'' and ``many'' is a straightforward
matter of logic. And is therefore unexceptional. (Unconditional).
But we have already seen in our definition that not all kinds of
Incompatibility are unexceptional; that is to say, [b]. How then
does the incompatibility expressed by [b] come about?

Here is an example: I have a daughter and, besides, 10.000 dollars
in the bank. No problem there. But then my daughter is kidnapped
and I receive a ransom note for 10.000 dollars. I can no longer
have both, my daughter \emph{and} 10.000 dollars in the bank. In
view of the specific circumstances confronting me, a pair of
hitherto compatible situations have been rendered ``mutually
exclusive'', Bohr's known term for inaugurating his introduction
to Complementarity (CTY hereafter). ``Having one's daughter'' and
``having 10.000 dollars in the bank'' are not incompatible states
\emph{per se} in the least. But they can be made incompatible,
provided that the right sort of suitable \emph{conditions} are
introduced. In their face any two, hitherto compatible states (or
concepts) can be rendered incompatible, \emph{on condition} that a
suitable set of physical conditions are obtaining or provisionally
introduced, thereby forbidding their hitherto recorded mutual
compatibility for the entire duration of their presence. It goes
without saying that, once the presence of such conditions is
removed, the two (temporarily) disjunctive states will become
mutually compatible once again in their usual, peaceful
coexistence.

We have seen, therefore, that in the case of conditional
incompatibility between a pair of states (i.e. of incompatibility
type [b]), it takes the intervention of an \emph{additional fact},
if it is to ever result. This we may entitle ``\emph{the
prohibitive fact}''. (\cite{AN94}, p. 188). It should be stressed
that the ``prohibitive'' element in question is invariably and
uniquely an additional \emph{fact} and nothing over and above a
(mere) fact. And as such, unexpected from a formal point of view.
By contrast, unconditional incompatibility should never be
unexpected from a formal point of view, because it is a matter of
\emph{logic}. Not a matter of fact which could, formally at least,
have gone the other way. The two pairs of antithetic clusters,
``factual/conditional''-``logical/unconditional'', are therefore
individually coextensive, respectively.

Conditional incompatibility cannot result by virtue of the
\emph{definitions} of the (currently) incompatible pair of states
(or pair of concepts). If it could, their incompatibility would be
logical and, as such, unexceptional; in other words,
unconditional. It therefore follows that all instances of
conditional incompatibility will invariably turn out to be
factual, whereas, by evident contrast, all instances of
unconditional incompatibility invariably logical. From this
realization follows a further consequence, the importance of which
to the overall argument we can hardly overemphasize. The preceding
considerations have unequivocally established that
\emph{unconditional incompatibility is \textbf{self sufficient}}.
By contrast, once again, conditional incompatibility is
\emph{never} self sufficient. It invariably requires an
\emph{additional}, prohibitive factor, capable of driving the two
thus related states to incompatibility, an incompatibility which
otherwise -and in absence of the said factor - would itself be
impossible to result.

The two formal (and exhaustive\footnote{Exhaustive, that is to
say, in Two-Valued Propositional Calculus.}) definitions of
incompatibility previously specified, i. e. those of
incompatibility type [a] and incompatibility type [b], succeed in
reflecting the property of Self-Sufficiency -or its absence- quite
explicitly. In formula [a] the incompatibility is \emph{confined}
to the two related variables, $p$ and $q$, at the exclusion of all
other conditions, and we are forbidden to go looking beyond the
two variables \emph{per se} for its establishment. In fact, to go
looking \emph{beyond} the two variables of relation $(p
\rightarrow\neg q) \wedge (q \rightarrow\neg p)$ for tracing or
grounding their incompatibility is, quite simply, contradictory to
the assumption. If only to repeat the point, unconditional
incompatibility is (\emph{intolerantly}) self sufficient.

In formula [b], however, the situation is altogether different.
The biconditional connective, $\Leftrightarrow$, speaks for
itself. The variables $p$ and $q$ will never in the context of
formula [b] enter a relation of mutual incompatibility without
help from outside. $p$ and $q$ will simply be compatible without
such help, as can be seen immediately from assuming  $\neg r$. The
outside help is withdrawn and the variables become compatible.
Consequently, on the whole, unconditional incompatibility is
synonymous with self-sufficient incompatibility and, accordingly,
conditional incompatibility synonymous with
self-\emph{insufficient} incompatibility. Emphasis on this
provision, though it may seem pedantic to most readers at this
stage, is nonetheless well warranted and many quantum surprises
will depend on it.

Now that the distinction between conditional and unconditional
Incompatibility has been defined and understood as above it is
time to turn and ask the next, natural question: Do the two
quantum uncertainties, $\Delta E \Delta t \geq h$, $\Delta p
\Delta q \geq h$ express conditional or do they express
unconditional incompatibility between their two related sets of
variables, $E$ with $t$ and $p$ with $q$? In view of the preceding
reasoning the answer to this question is similarly natural. The
reciprocal uncertainties in the values of the two pairs of
conjugated variables, $E$ with $t$ and $p$ with $q$, as presently
joined, obviously express \emph{conditional} incompatibility
between these variables. Conditional (obviously) on $h$ itself.
Clearly, for $h=0$ both clusters of related uncertainties would
vanish. On the other hand, they do emerge for $h>0$. Consequently,
$\Delta E \Delta t$, $\Delta p \Delta q$ are uncertainties which
are there because and \emph{only} because of $h$. And would be
removed in its absence. This reads, respectively,
                                        \begin{center}$\Delta E \Delta t\Leftrightarrow h>0$ and, accordingly, $\Delta p \Delta q\Leftrightarrow
                                        h>0$\end{center}
which both precisely correspond to logical formula [b]. Evidently,
then, the two UR express conditional incompatibility between their
related variables, conditional, that is, on nothing other than
$h$.

The interpretation thus suggested can then be integrated just as
naturally as all the other elements so far were in the following
(natural) fashion: The two pairs of conjugate classical variables,
$E$ and $t$, $p$ and $q$, yielding the two action products Et and
pq of the corresponding uncertainties, are rendered incompatible
in QM because, simply, the latter theory incorporates an
\emph{additional fact}, hitherto unacknowledged and unanticipated
by the classical theory, namely, action quantization, and it is
the intervention of this precise fact, absent in classical
assumptions, which is responsible for the incompatiblity in the
joint determinations of $E$ and/or $t$ and $p$ and/or $q$ below
its limit, $h$. The two sets of incompatibilities are therefore
\emph{fact dependent}, that is to say, conditional on a
\emph{fact}; $h$. And therefore, trivially, express conditional
incompatibility only. One of the most reliable commentators in the
field, C.A.Hooker, certainly seems to think so and not at all
without reason:
\begin{quote}
Bohr believes  that while it has seemed to us at the macro-level
of classical physics that the conditions were in general satisfied
for the joint applicability of all classical concepts,  we have
\emph{discovered}  this century  that this is not accurate and
that the conditions required for the  applicability of some
classical concepts are actually incompatible with those required
for the  applicability of other classical concepts. This is the
burden of the doctrine (B4) [=CTY.]

This conclusion is  necessitated by  the discovery of  the quantum
of action \emph{and only because of its existence}. It is not
therefore a purely \emph{conceptual} discovery that could  have
been made  \emph{a priori} through a  more critical
\emph{analysis} of classical concepts. It is  a discovery of the
\textbf{factual absence} of the conditions required for  the joint
applicability of certain classical concepts.[\cite{HOO} Dark
letters for the author's italics.]
\end{quote}

This, therefore, is exactly as foretold. The incompatibility above
referred to is factual, because it is not the product of concept
analysis, disclosing a logical discrepancy between the disjunctive
concepts (and as such available a priori) and, therefore, as being
fact \emph{dependent}, it is eo ipso conditional. Conditional,
that is, on the fact itself upon which it is dependent, and which
we have previously labelled ``the prohibitive fact''. In other
words, the quantum. Perhaps a fleeting allusion to the spontaneity
of the author's account of the matter, its `naturaleness' so to
speak, would not be entirely out of place. Spontaneity is
important in this instance because it serves in crosschecking the
two accounts, ours, which is in conscious awareness of a contrast
between these two types of incompatibility, formally defined, and
Hooker's, which is rather intuitive and reflexive at this
stage\footnote{And at a subsequent stage he has repudiated it
altogether! Indeed, at a later time Hooker has actually expressed
his \emph{scepticism} as concerns the viability of the distinction
between logical and factual incompatibility as formulated
by~\cite{AN94} or, even, its usefulness as such. In his letter to
Antonopoulos, dated 18 December 1989, he says that when it comes
to ``formal'' as opposed to ``factual'' aspects of the problem at
hand ``naturalists like me [him] cannot make a sharp distinction
between the two kinds of truth''. Well, up \emph{there} he has!
Not too consciously, it would appear, but nonetheless most
definitely. Which is all to the better, really, for, when not too
much undue sophistication has come by just yet, to hold one
captive to wavering amphiboly, first thoughts are best thoughts.}.

On the whole, therefore, at first it seems a safe bet that the two
pairs of classical variables of QM, when featuring pairwise in the
two corresponding quantum uncertainties, they should express
conditional incompatibility between the thus related concepts and
nothing but that. On closer inspection, however, the situation
appears a great deal more complex than initially assumed. Closer
inspection in fact reveals that, when analyzed and examined all
across the logico-mathematical board, the quantum uncertainties
manifest and force upon us an incompatibility which is
\emph{both}; conditional \emph{and} unconditional for one and the
same pair of classical concepts.

Amazingly, the same phenomenon is noticeable, as we shall
demonstrate later on, right within the noncommutative formalism
itself. To the fundamental, noncommutative formalism inaugurated
by the formula $pq-qp=i\hbar$ there is now erected the
noncommutative formalism of $AB \neq BA$. In other words, a
noncommutativity without the quantum! There is hardly ever a
commentator who would not treat the two commutativities as
interchangeable, with the sole exception, in our knowledge, of
Hilgevoord and Uffink~\cite{HU}. But they are not really
interchangeable at all. One is the \emph{mathematical} consequence
of non-diagonal matrix multiplication, yielding noncommutativity
by definition, the other a noncommutativity \emph{due to $\hbar$}.
There's a difference.

\section{Applying the Distinction}

\subsection{Application to Wave-Particle Duality}
The results of applying the Conditional vs Unconditional
Incompatibility distinction to quantum problems appear quite
startling when viewed in this light. As a rule observed by nearly
all physicists, the quantum uncertainties and Wave-Particle
Duality (WPD hereafter) are treated as if intimately associated.
And, indeed, there is a strong temptation to associate them.
According to this association a certain group of classical
variables by nature relate to the particle, their complementary
variables by nature to the wave. But particles are local entities,
so particles are small. By contrast, waves are nonlocal entities,
so waves are large. And the opposition between large and small is
logical, that is to say, fact \emph{in}dependent. Hence, waves and
particles are \emph{self-sufficiently} incompatible. This is why,
besides, waves and particles are incompatible \emph{\textbf{also
in classical mechanics}}. And classical mechanics does not contain
the quantum.

Well, then. If the two UR are a consequence of WPD, one set of
variables belonging to the wave, the other set to the particle,
then, since waves (large) and particles (small) exclude one
another \emph{self-sufficiently}, and hence without any help from
the quantum, the variables appearing in the two UR, as derived
from WPD, would \emph{also} exclude one another self-sufficiently,
and hence exclude one another without any help from the quantum.
In fact, they do not need any help from anything at all, except of
course the \emph{self contained} opposition between ``large'' and
``small'' itself. Which opposition, as remarked, obtains
independently of the quantum. Consequently, either the two UR have
nothing to do with WPD or else they have to do with
WPD\footnote{Which is what one of us has been insisting for two
decades now; See \cite{AN88} and, especially, \cite{AN94}.}, but
then they have nothing to do with the ...\emph{quantum}, on which,
however, they are supposed to depend!

In other words, how can the incompatibility contained in WPD,
which is self sufficient enough to obtain full force even in
\emph{classical mechanics}, ever be responsible for the
incompatibility between the classical conjugate variables, which
latter results only on the basis of \emph{quantum} assumptions?
Or, to put the point differently, how can a fact independent
incompatibility, as that belonging to WPD, ever be responsible for
a fact dependent incompatibility, as that demanded by the two
quantum uncertainties?

Some people still believe that WPD is the epitome of the quantum
uncertainties, if not indeed the epitome of QM as such. But once
the Conditional/Unconditional Incompatibility contrast is applied
to it, it simply proves to be an incoherence. The uncertainties,
exactly as Hooker stressed, must absolutely depend on the quantum
or be nothing at all. But if the uncertainties are constructed
upon the logical model afforded by WPD, they will thereby express
a \emph{self-sufficient} type of incompatibility and, as we have
seen, such incompatibility -trivially- has no need of the quantum.
To be precise, cannot even make room for the quantum, except
contradictorily. People think that WPD furnishes the right sort of
quantum incompatibility required by the UR. We have just shown
that it furnishes the wrong sort, if there ever was one.
\subsection{Application to $\Delta E \Delta t \geq h$}
But the real trouble does not lie in the comparison between an
invalid derivation and a -let us say- valid one. The invalid one
can be discrarded at no cost. The real trouble lies within the
frame of the valid derivation itself. For that too is equally open
to both accounts, the conditional \emph{and} the unconditional.
Consider the Fourier reasoning applied to the quantum relation
$E=h\nu$. (See~\cite{HU} and Bohr's own work referred to there;
for more detail see Marmet, 1994, p.343; see,
finally,~\cite{AN88})

Fourier's known relation, $\Delta\nu\Delta t \geq \frac{1}{2\pi}$,
was based on the observation that it is a logical impossibility to
determine the frequency at an instand $dt=0$. Frequency is by
definition a repetitive phenomenon and hence by definition such as
requires a time latitude to be exemplified, if at all. Obviously,
I cannot define the regular reoccurrence of a certain event over
even time intervals within a time $dt=0$, i.e. a time so narrow
that won't allow the event to occur even once. As D.M. Mackay has
remarked almost fifty years ago, the idea of defining a frequency
at an instant $dt\rightarrow 0$ is self contradictory. ``This is
not physics but logic'', he says~\cite{MA}.

Once the quantum relation, $E=h\nu$ is (factually!) established,
by simply substituting for $\nu =\frac{E}{h}$ in Fourier's above
mentioned relation, we immediately obtain $\frac{\Delta E}{h}
\Delta t \geq \frac{1}{2\pi}$ and, finally, $\Delta E \Delta t
\geq h$. Now, what \emph{sort} of incompatibility does  express,
if derived in this way? Well, it should express precisely the sort
of incompatibility which $\nu$ itself, the frequency, does in
Fourier's relation. Are we not constantly reminded that ``energy''
is the frequency in QM? Mackay, for instance, speaks of
\begin{quote}
the \emph{identification} of energy with frequency~\cite{MA}.
\end{quote}
and, in much more recent times, we are told in no uncertain terms
that
\begin{quote}
This simple Planck relationship between energy and light frequency
in effect says that energy and frequency are \emph{the  same
thing}, measured in different units~\cite{CH}.
\end{quote}
A shorthand expression of the whole idea is the relation
$E\approx\nu$, which says it all. So to the task of specifying the
syllogistic mechanism involved:
\begin{itemize}
\item Premise 1: Energy is logically equivalent with the frequency. \item Premise 2: Frequency is logically incompatible with an exact time.
\item Conclusion: Hence, \emph{energy} is logically incompatible with an exact time.
\end{itemize}

Here \emph{Conclusion} follows from Premises 1 and 2 as trivially
as the proverbial mortality of Socrates follows from ``All Men Are
Mortal'' and ``Socrates Is A Man''. Attention should be paid to
the subordinate predicate, ``logically'', modally conditioning the
primary predicate, ``incompatible''. If Energy is coextensional
with the Frequency, this predicate must necessarily be included in
the \emph{Conclusion}, otherwise, and in its absence, the latter
will not be validly drawn, contradicting their coextensionality.
Hence, in the most straightforward and valid of manners, energy is
above shown to be \emph{logically} incompatible with an exact
time, just as frequency previously was. But, as we have seen,
concepts incompatible in this sense are self - sufficiently so.
And concepts which are self - sufficiently incompatible are
concepts whose incompatibility is in fact independent. And
therefore such that cannot even \emph{relate} to a fact, e.g. $h$.
Hence, in accordance with the Fourier treatment of the relation
$E=h\nu$, we obtain an uncertainty $\Delta E\Delta t$ due to a
fact, $h$, with which it cannot even relate.

The reactions to this conclusion are not too difficult to foresee.
Fortunately, we have at our disposal something a good deal more
substantial than mere foresight to get our hands on, namely, an
actual objection recently raised. It is the following:
\begin{quote}
I cannot share the author's diagnosis. The energy - time
uncertainty relation can be derived from two premises: (1)$E=h\nu$
(2)$\Delta \nu\Delta t\geq 1$. Here it is clear that the second
relation is the result of Fourier analysis, and therefore
independent of any physical assumption. The first however is
clearly a non-trivial \emph{physical} assumption, that need not
hold in physical theories other than QM. (1) and (2) together
imply $\Delta E\Delta t\geq h$ (3). The diagnosis is simply this:
since conclusion (3) depends on two premises, one of which is
dependent upon a physical assumption, the conclusion\footnote{The
conclusion should be dependent on this physical assumption, $h$ or
$E=h\nu$, says the author of the passage, correcting us. But we
have never denied that it \underline{depends} on this assumption.
Anything but. We have only raised the question, whether the
physical assumption referred to RETAINS ITS IDENTITY. We have
never denied \underline{whether} $E=h\nu$ is a premise to the
argument. This is precisely what we have stressed. We have only
questioned the NATURE of this premise and whether the CONTEXT of
the argument, imposing the logical relation $\Delta \nu\Delta
t\geq 1$, allows it to reatin its original logical properties or
whether it retrodictively cancels them,given the overall pressures
of the said context.} is dependent on this [physical] assumption
too.\footnote{Extract from a report on a previous version of this
paper, dated 4th November 2003. Italics, brackets and initial
ours. The report was written for \emph{Studies in the History and
Philosophy of Modern Physics} and is at the disposal of the Editor
of.}
\end{quote}
And hence must be dependent on $h$. This is all so nice and cozy
and so consonant with quantum tradition that hardly anyone would
resist the temptation of replying just thus, a referee all the
more so. However, it takes but one word to spoil it all, its hopes
and plans included, though not necessarily the fun as well:
Substitution. Once this word is properly attended to, this
objection is exposed in all its circular and dogmatic
incorrigibility.

What is the true essence of the entire Fourier derivation? It is,
in a word, the \emph{\textbf{substitution}} of $\nu$ in $\Delta
\nu\Delta t\geq 1$ by $E/h$ in order to derive $\Delta E\Delta
t\geq h$. And in order that one can be at all entitled to
substitute $E/h$ for $\nu$ one needs to presuppose that the two of
them, the substitute and the substituted, will just have to be
identical, or equal, or equivalent or what have you, provided they
are so intimately linked as to license and, indeed, entail the
substitution. You name it, they have to be it. In consequence,
$E/h$, which replaces $\nu$, the frequency, \emph{is} the
frequency, or else the substitution is illegitimate and has no
business being there in the first place. And then, since $E/h$ is
the frequency, what is true of the frequency must be true of its
substitute, $E/h$. And then, since what is true of the frequency
is that it is unconditionally incompatible with time, which this
referee openly concedes, $E/h$ is also unconditionally
incompatible with time, which he inconsistently does not. It is
either that or else the substitution is sheer bogus and no $\Delta
E\Delta t\geq h$ of any kind will result, coherent or otherwise.

By right of mathematical law, the law of substitutions,
$E/h(=\nu)$ is unconditionally incompatible with time, \emph{even}
if it deceitfully contains $h$ in the denominator of the fraction
just to mislead(some of) us. The conclusion can now be denied at
the pain of contradiction. By inserting $E/h$ in the place of
$\nu$ in $\Delta \nu\Delta t\geq 1$, we commit ourselves to making
$E/h$ whatever $\nu$ is, thus deriving a logical uncertainty and,
therefore, a fact independent one that cannot even relate to this
very $h$, which it has itself put there! In the face of our
distinction the Fourier tratment of $E=h\nu$ leads to incoherence
and absurdity comparable to that of WPD previously
encountered(essentially it is the same exact problem), if not
indeed to a worse kind. Valid reasoning is reasoning which
transmits the logical properties of the premises down to the last
conclusion. And the logical properties of premise $\Delta
\nu\Delta t\geq 1$ is that it incorporates a self - sufficient
type of incompatibility, which renders $h$ redundant.

The essence of the problem here encountered stems from the fact
that, in view of the distinction here introduced (and hitherto
absent in all quantum theorizing), $E=h\nu$ proves a full scale
logical \emph{hybrid}. Initially, $E=h\nu$ states a factual truth
-a startling one at that- so whatever $E$, $t$ incompatibility is
subsequently destined to result on its basis, it should only be
fact-dependent in this particular context. However, what this
(unique) factual truth reveals right after is, that the concept
which is (factually) equivalent with $E$, i.e. the frequency,
$\nu$, is itself \emph{logically} incompatible with an exact t,
thereby rendering the thus resulting $E$, $t$ incompatibility a
fact independent one, in this other context. Given the
surrounding, `outer' context, i.e. the factualness of $E=h\nu$, E
and t must be conditionally incompatible. But given the
\emph{surrounded}, `inner' context, i.e. the logical
incompatibility between a $\Delta\nu=0$ with a $\Delta t=0$, E and
t must now be unconditionally incompatible.

When, in other words, $E=h\nu$ is considered in its (outward)
relation to reality, it must in this capacity be a factual truth.
But when considered in itself (inwardly), in this other capacity
it incorporates a logical truth. What should we say then? That
what $E=h\nu$ really asserts is that, on its basis, $E$ and $t$
are unconditionally incompatible concepts \emph{\textbf{on
condition}} that $E=h\nu$ is true? On the basis of the distinction
here introduced this is exactly what we have to say. Though, of
course, in its absence, we wouldn't have to.

\section{The hybrid nature of Quantum Mechanical Formalism}
In Heisenberg's paper of 1925\cite{HEI} there is mentioned a type
of multiplication between the quantities characteristic of a
quantum system directly leading to relations of noncommutativity
between them. Such multiplication was subsequently identified by
Born and Jordan\cite{BJ} as a multiplication of matrices
corresponding to the physical quantities attributable to a quantum
system. This was the inauguration of transformation theory which
in turn developed into the widely disseminated axiomatic
foundation of von Neumann's\cite{VN}.

In the following pages of the present essay we shall attempt to
classify Heisenberg's quantum multiplication -this is how it will
be referred to from now on so that it will be distinguished from
matrix multiplication as such-, the multiplication of matrices and
their concomitant noncommutativities, and finally the resulting
uncertainties, on the basis of  the distinction established in the
first part of the paper. In particular, the relations mentioned in
Heisenberg's paper are satisfied by physical systems \textbf{on
condition that action is quantized and on that condition alone}.
By contrast, the mathematical treatment, which was initiated by
Born and Jordan, constitutes a \textbf{``hybrid''} for the second
time running, because the premises of this latter hypothesis may
lead to the noncommutative relation $pq-qp\neq 0$ for the
variables $p$,$q$, but the specific relation $pq-qp=i\hbar
I$\footnote{This formula is refered to as canonical commutation
relation. In \cite{HEI} there is only a specific form of this
relation.} is not intrinscically derivable from within it. It is
extrinscically introduced on the basis of further assumptions.

In other words, the latter noncommutativity is inherent in advance
within the chosen formalism, as a self-subsisting mathematical
property, contrary to Heisenberg's multiplication, which is
factually \emph{dependent} on the quantum and cannot result in its
absence. Whereupon, the noncommutativity in question must become
system-specific in order to be applied to the relevant phenomena.

Commencing, Heisenberg dennounces the classical picture of an
electron's kinematics and proceeds to the adoption of a different
interpretation of the function $x(t)$ whose classical
interpretation would be the particle's position in the space of
intuition. In a parallel course, however, considering the
correspondence principle, he retains the differential equation
which governs the said function (Newton's second law) in the
classical treatment. Thus he accepts that the equation
$\ddot{x}(t)+f(x)=0$ regulates the connection of $x(t)$ with the
outwardly exerted force $f(x)$.

In what follows he analyzes $x(t)$ in ``Fourier'' fashion, so that
the resulting expression will harmonize itself with the quantum
conditions. In the classical case, if $\nu(n,a)$ is the frequency
observed during the transition from state $n$ to state $a$, then
$\nu(n,a)=\frac{a}{\hbar}\frac{dW(n)}{dn}$(1) Where $W(n)$ is the
energy of the said state. By contrast, due to the presence of
discontinuous states in the quantum case, the frequency during the
transition from state $W(n)$ to state $W(n-a)$ is characterized by
emission of radiation $\nu(n,n-a)=\frac{W(n)-W(n-a)}{\hbar}$(2).
Suppose then that $x(n,t)$ is $x(t)$ in the specific case that the
electron is in the state $W(n)$. Then, classically, $x(n,t)$ would
be expanded as
$\int_{-\infty}^{+\infty}U_{a}(n)e^{i\omega(n)at}da$(3), where
$U_{a}(n)$ is now a complex quantity, i.e. the transition
amplitude, whose squared measure furnishes the probability that an
electron will pass from state $W(n)$ to state $W(a)$. Quantum
mechanically, $x(n,t)$ is expanded into a series
$\sum_{a=-\infty}^{+\infty}U(n,n-a)e^{i\omega(n,n-a)t}(4)$\footnote{With
the assumption that the system is periodic or multiperiodic. Else
the series has to be replaced by the integral
$\int_{-\infty}^{+\infty}U_{a}(n)e^{i\omega(n,n-a)t}da$ without
any essential change in the foregoing argument.}, where $U(n,n-a)$
plays a part analogous with $U_{a}(n)$ and
$\omega(n,n-a)=2\pi\nu(n,n-a)$. Then in accordance with the form
assumed by $f(x)$ there are obtained retrodictive formulae for the
quantities $A(n,n-a)$ and $\omega(n,n-a)$, where
$A(n,n-a)=Re\{U(n,n-a)\}$, introducing the expansion into the
differential equation.

Consider then two quantities\\
     $\alpha(t)=\sum_{a=-\infty}^{+\infty}U(n,n-a)e^{i\omega(n,n-a)t},\beta(t)=\sum_{a=-\infty}^{+\infty}V(n,n-a)e^{i\omega(n,n-a)t}$. Then
     $\alpha(t)\beta(t)=\sum_{b=-\infty}^{+\infty}Z_{1}(n,n-b)e^{i\omega(n,n-b)t}$ and\\
     $\beta(t)\alpha(t)=\sum_{b=-\infty}^{+\infty}Z_{2}(n,n-b)e^{i\omega(n,n-b)t}$,\\where
     $Z_{1}(n,n-b)=\sum_{a=-\infty}^{+\infty}U(n,n-a)V(n-a,n-b)e^{i\omega(n,n-b)t}$ and
     $Z_{2}(n,n-b)=\sum_{a=-\infty}^{+\infty}V(n,n-a)U(n-a,n-b)e^{i\omega(n,n-b)t}$.

Whereupon, in general, we obtain
$\alpha(t)\beta(t)-\beta(t)\alpha(t)\neq 0$. That is to say, the
multiplication of the two quantities ceases being commutative. And
this noncommutativity exists by virtue of the quantum of action.
Were it not for the quantum, we would not have observed a
discontinuous and countable sequence of states, starting from the
ground state. In consequence, it would not be formula of minuses
nr.(2), which would then obtain, but nr.(1). But then, as can be
verified by a single calculation,
$\alpha(t)\beta(t)-\beta(t)\alpha(t)=0$. This result is directly
specified for the magnitudes $x(t)$, of position, and
$p=m\dot{x}(t)$, of momentum.

We conclude by contending that the noncommutative quantum
multiplication here obtained is satisfied only on the basis that
$h>0$ and would reduce to ordinary, commutative multiplication,
were the limitation $h>0$ to be withdrawn. As will soon become
evident this is no longer true for transformation theory and, by
extension, for von Neumann's axiomatization. Noncommutativity is
still there but it is now of a different type, resulting as a self
contained property of the mathematical scheme employed. And which
is therefore unconditionally true, i.e. such that obtains
independently of $h$. To compensate for this, it then has to be
remodified in order to accord itself with the phenomenon
investigated, and thus it is inconsistently reshaped into a
noncommutativity by virtue of facts on top of the previous. This
may not suffice to condemn all parts of the formalism, but it can
be a severe problem.

Commencing their treatment Born and Jordan assume that the
conjugate dynamical variables of the system under investigation,
$p$,$q$, constitute hermitian matrices of the form
$p=(p(nm)e^{i\omega(nm)t})$ and $q=(q(nm)e^{i\omega(nm)t})$, where
$m,n\in\mathbb{N}$,
$\omega(nm)=\omega(n,m)=2\pi\frac{W(n)-W(m)}{\hbar}$ in
Heisenberg's symbolism and $q(nm),p(nm)\in\mathbb{C}$ generally.
Due to this the products $pq$ and $qp$ are considered to be the
products of a pair of matrices. It is commonplace knowledge that
matrix multiplication is the archetype of noncommutative
multiplication in Mathematics. Therefore, generally $pq-qp \neq 0$
(5) without the quantum playing a part in this effect. What is
more,  it is shown next  in the work referred to  that the
relation $pq-qp=i\hbar I$, where $I$ is the identity matrix, by
means of Bohr's quantum formula $J=\int_{0}^{\frac{1}{\nu}}pdq=
nh$. It is more than evident that this last formula constitutes
the premise, so as to make matrix mechanics agree with the
phenomenon, which in Heisenberg's argument has perfectly sufficed
for deriving the quantal multiplication without any external help.
The authors of the paper themselves remark that relation (5)
constitutes a direct statement of the correspondence principle.
That is to say, when $\hbar\rightarrow 0$, $pq-qp\rightarrow 0$,
and the quantities become commutative. However, this statement is
misleading.

It is true that $pq-qp\rightarrow 0$, when $\hbar\rightarrow 0$,
but this ``0'' is the zero matrix and not the zero of real
numbers. If that relation constituted a genuine statement of the
correspondence principle, $p$ and $q$ would in the end turn out
real numbers, and this is not what happens. Quite simply, from
noncommutative matrices they turn into commutative ones. This will
not of course degenerate all the eigenvalues of each and every
matrix into one, and so the quantum fluctuations will remain
invariant. The difference is that the conjugate variables coevolve
without excluding one another. Although the two multiplications,
the quantal and the matrix, display the same formal attitude, this
will not coerce their referents to become identical.

By contrast, in Heisenberg's reasoning, when $\hbar\rightarrow 0$,
the ``Fourier'' series assumes a classical expansion, since we can
now differentiate by using formula (1) and the two quantities turn
out to be real numbers. In this way, while the two
multiplications, the quantal and the matrix, manifest the same
outward effects, the mathematical objects which each one refers
are quite distinct.

And it is at this point that the inconsistent deviation from
Heisenberg's multiplication is inaugurated. Conflating the two
cases, as we have seen it done before (1st Part) Born, Jordan and
von Neumann proceed on the supposition that the two
noncommutativities were analogous or identical and, hence, that
the two formalisms were.    But they are not. For while in
Heisenberg's multiplication, on the assumption that ``$h=0$''
noncommutativity is eliminated, in the case of \cite{BJ} where a
zero matrix is obtained instead, noncommutativity is not. It just
assumes a different look by putting on a mask, yet without ever
departing from its authentic identity.

On our understanding of the matter, the problem is rather simple
to state though not necessarily simple to solve. The
noncommutativity in Heisenberg's multiplication is eliminable in
principle because it is conditional on the quantum, hence
removable in its absence, but the noncommutativity of Born, Jordan
and von Neumann is ineliminable in principle, because it is self
subsisting. It is a fact-independent noncommutativity and
therefore ineliminable come what may. It is a noncommutativity
ascertainable in advance, a clear cut mathematical phenomenon the
residues of which stay on in one form or another, once this
mathematics is employed. Not being conditional on anything except
its own self it therefore continues to tacitly obtain even when
the quantum is removed, yielding for classical expectations a zero
matrix only, instead of zero just, bearing witness to its own
fact-independent origin.

Consequently, it will either be consistently regarded in its pure
mathematical essence, whereupon however it cannot even relate to
the quantum, or else incorporate Heisenberg's conditional
noncommutativity, if to be at all able to apply to specific
quantum problems, but then do so at the price of an antinomy. The
very antinomy detected in our first part of the argument, where
the logical $\Delta \nu \Delta t \geq 1$ is turned into the
factual $\Delta E \Delta t \geq h$, when the empirical relation
$E=h\nu$ is introduced. Yielding in both instances the same
logical hybrid.

Before we conclude we wish to make explicit the situation in von
Neumann's abstract axiomatization\footnote{We shall study
autonomous(isolated) systems. So the Hamiltonian is
time-independent and corresponds to the totel energy of the
system}. Let us list the axioms in this approach:
\begin{itemize}
\item[1.]To each quantum system there corresponds a complex separable Hilbert space $(\mathcal{H},\langle\cdot,\cdot\rangle)$
    where $\langle\cdot,\cdot\rangle$ is the inner product. Every $\psi\in\mathcal{H}$ with $||\psi||=\sqrt{\langle\psi,\psi\rangle}=1$ corresponds to a state of the system.
    By equivalence, the projectile space $P\mathcal{H}=\{[\psi]:(\psi\in\mathcal{H})\wedge(||\psi||=1)\}$ constitutes the set of
    the states of the system.
\item[2.] The observables of the system in question are symmetrical operators of space $(\mathcal{H},\langle\cdot,\cdot\rangle)$
    that is to say, observable $A$ is represented by a linear operator $A$ such that $\langle
    A\phi,\psi\rangle=\langle\phi,A\psi\rangle,\forall\phi,\psi\in\mathcal{H}$. The values of the observable $A$ are the spectrum of the
    corresponding symmetrical operator, $\sigma(A)=\{\lambda\in\mathbb{C}:\textrm{$A-\lambda Id_{\mathcal{H}}$ is singular}\}$\footnote{Cause of the operator's symmetry the spectrum is a subset of the real numbers}, where
    $Id_{\mathcal{H}}$ is the identity mapping. The mean value of the observable, $E(A)$, is given by the corresponing spectral measure,
    $E(A)=\langle\psi,A\psi\rangle$, where $\psi$ the state of the system.
\item[3.] The Hamiltonian of the system, $H$, constitutes an observable and its corresponding
    operator is self-adjoint. In Heisenberg's representation, if $A$ is an observable, then
    its temporal evolution is determined by the equation $\dot{A}=\frac{i}{\hbar}[H,A]$, where $[H,A]=HA-AH$ is the
    commutator of $H$ and $A$. In Schroedinger's picture, when the system is in state $\psi(t_{0})$
    at the time $t_{0}$, it is in the state $\psi(t)=exp(-\frac{i}{\hbar}H(t-t_{0}))\psi(t_{0})$\footnote{We `ve used here the Stone theorem, see \cite{ST}}(formally the Schroedinger equation is
    then written as  $i\hbar\frac{\partial\psi}{\partial t}=H\psi$). In every one of the cases the Hamiltonian constitutes an
    infinitesimal generator of the evolution of the dynamical quantities.
\end{itemize}
The first two axioms introduce a self-contained noncommutativity,
since the observables are represented as symmetric operators in a
Hilbert space. The sole case, when the symmetric operators always
commute, is that of a one-dimensional Hilbert space, that is to
say $\mathcal{H}=\mathbb{C}$. But then the system should have only
one state which could occupy. It is evident why we can't accept
this strongly artificial situation. If we demand, however, of the
quantities of momentum and position to fulfill the normal rules of
commutativity, the resulting space must now be one of infinite
diamensionality, as Born and Jordan themselves remark. (To be
precise, the operators of momentumn and position cannot be bounded
\cite{WIE,WI}, and this is why in the foregoing axioms we speak of
symmetric rather than of self-adjoint operators.)

Consequently, noncommutativity is in this case inherent within the
axiomatic system a priori, as is inherent in the statetement, ``If
$X>10$, then $X>5$'', the statement that ``then $X>6$''. Which
latter is a trivial consequence of the previous, hypothetical
statement. Similarly the noncommutativity in question is a
tautological consequence of the axioms 1 and 2 and has nothing to
do with the quantum of action. Further ahead, in the third axiom
the quantum emerges in the evolution equation of each picture.
This, however, constitutes a further introduction for settling the
matter, exactly as it has in the papers of Born and Jordan, and
that of Born, Jordan and Heisenberg which followed.

Operating within the frame of von Neumann's axiomatization we can
demonstrate Robertson's general uncertainty relation: Let $A$,$B$
be two observables. Then the inequality
$(Var(A)Var(B))^{\frac{1}{2}}\geq\frac{1}{2}|E([A,B])|$ obtains,
where $Var(A)$ the variance of A, $Var(A)=E((A-E(A))^{2})$(the
same for $Var(B)$).

This inequality is also self-sufficient and obtains without
mediation of the quantum. If we consider the relation
$[p,q]=i\hbar Id_{\mathcal{H}}$ for position and its conjugated
momentum then $(Var(p)Var(q))^{\frac{1}{2}}\geq\frac{\hbar}{2}$ or
equivalent $\Delta p \Delta q\geq \frac{\hbar}{2}$, if $\Delta
A=\sqrt{Var(A)}$ is the uncertainty dispersion of quantity $A$,
that is to say, Heisenberg's uncertainty relation for momentum and
position. But this uncertainty should be dependent on the quantum
and still its derivation, down to its terminal conclusion, has
been quite possible without it.

We therefore notice that although Robertson's inequality reflects
a self-sufficient noncummutativity and hence independent of $h$,
by introducing further premises, we end up with a hybrid
noncommutativity, due to and not due to $h$ in the end. In other
words an uncertainty relation that is fact dependent and fact
independent at the same time. Higevoord and Uffing argue\cite{HU}
that Robertson's inequaltity presents problems additional to the
one we have detected.

In concluding we contend that the mathematical formalism has since
1925 been tracking a most mysterious and confusing path. On the
one hand it imposes upon the physical magnitudes involved
noncommutativities warranted \emph{a priori}, and hence such as
were true in advance of the factual discovery that $h>0$, and on
the other hand introduces a different kind of noncommutativity
altogether, in order to force the former to come to agreement with
the physical content it purports to reflect. We are not pursuing
the deeper causes or motives behind this contradictory tendency
but we do believe that steps towards its amendment should be
taken, so that the formalism may maintain its applicability and at
the same time restore the authenticity of the physical ideas which
have given rise to it.

\bibliographystyle{unsrt}
\bibliography{bib}
\end{document}